\documentclass[prb,aps,twocolumn,tightenlines,
showpacs,floatfix]{revtex4}
\usepackage{graphicx}
\usepackage{amssymb}
\usepackage{amsmath}

\begin{document}
\title{Intense terahertz laser fields on a two-dimensional hole
gas with Rashba spin-orbit coupling}

\author{Y.\ Zhou}
\affiliation{Hefei National Laboratory for Physical Sciences at
  Microscale and Department of Physics, University of Science and
  Technology of China, Hefei, Anhui, 230026, China}

\date{\today}
\begin{abstract}
We investigate the influence on the density of states and the density of
spin polarization for a two-dimensional hole gas with Rashba
spin-orbit coupling under intense terahertz laser fields. Via Floquet
theorem, we solve the time-dependent Schr\"{o}dinger equation and
calculate these densities. It is shown that a terahertz magnetic
moment can be induced for low hole concentration. Different from the
electron case, the induced magnetic moment is quite anisotropic due to
the anisotropic spin-orbit coupling. Both the amplitude and the
direction of the magnetic moment depend on the direction of the
terahertz field. We further point out that for high hole concentration,
the magnetic moment becomes very small due to the interference
caused by the momentum dependence of the spin-orbit coupling. 
This effect also appears in two-dimensional electron systems.
\end{abstract}

\maketitle

Optical properties of semiconductors are sensitive to
external conditions. Almost fifty years ago Franz\cite{Franz} and
Keldysh\cite{Keldysh} pointed out that under static electric
fields the absorption coefficient becomes finite below the band
gap, and the above-gap absorption spectrum shows oscillations. In
the late 1990s, Jauho and Johnsen \cite{Jauho_96,Johnsen_98}
studied the optical properties of semiconductors under strong ac field
and developed dynamic Franz-Keldysh effect (DFKE), which
presents the blueshift of the main absorption edge and the fine
structure near the band gap. This effect is particularly obvious for
semiconductors under intense terahertz (THz) field, thereby leading to
extensive theoretical and experimental interests on THz
electro-optics.\cite{Cerne_97,Kono_97,Nordstrom_98,Phillips_99} 
Very recently Cheng and Wu brought the spin degrees of freedom
into the study of the THz field induced effects. They studied a
two-dimensional electron gas (2DEG) with the Rahhba spin-orbit
coupling (SOC) under intense terahertz field.\cite{Cheng_APL_05} It is
shown that the THz field can efficiently modify the density of states
(DOS) of the 2DEG and excite a magnetic moment oscillating at THz
frequency. Later, Jiang et al. studied similar effects of quantum
dots\cite{Jiang_JAP_06}, and further discussed the spin dissipation
under THz driving fields.\cite{Jiang_PRB_07} However, all these works
concentrate on electron systems. Up to now there is no study on the
spin properties of hole systems under intense THz field. In this
letter we study the effect of intense terahertz laser fields on a
two-dimensional hole gas (2DHG) with Rashba SOC and show that this
system has some new properties different from the previously studied
electron system.

We consider a $p$-type GaAs (100) quantum well (QW). The growth
direction is denoted as the $z$ axis. A uniform THz radiation field (RF) 
${\bf E}_{\mathtt{THz}}(t)={\bf E}\cos(\Omega t)=(E_x,
E_y,0)\cos(\Omega t)$ is applied in the 
$x$-$y$ plane with the period $T_{0}=2\pi/\Omega$. The angle
between the electric field and $x$ axis is $\theta _{\bf E}$. By
using the Coulomb gauge, the vector and scalar potentials can be
written as ${\bf A}(t)=-{\bf E}\sin(\Omega t)/\Omega$ and
$\phi(t)=0$, respectively. We assume the well width is small
enough so that only the lowest subband is relevant. For this
structure, the lowest subband is heavy hole (HH) like. By applying a 
suitable strain, it can be light hole (LH) like. The confinement is
assumed to be large enough so that the lowest HH and LH subbands are
well separated and we can consider the HHs and LHs
separately. The Hamiltonian can be written as ($\hbar\equiv 1$) 
\begin{eqnarray}
  H_{\lambda}({\bf K},t)=\frac{{\bf K}^2}{2m^{\ast}_{\lambda}}
  +H_{so}^{\lambda}({\bf K},t)\ .
  \label{equ:H_tot_hole}
\end{eqnarray}
Here ${\lambda}=LH, HH$ and ${\bf K}={\bf k}-e{\bf A}(t)$, with
${\bf k}$ standing for the electron momentum. $m^{\ast}_{\lambda}$ is
the effective mass. For GaAs QW, $H_{so}$ is mainly due to
the Rashba term,\cite{Bychkov} and can be written as
$H_{so}^{\lambda}({\bf K})=\frac{1}{2}
(\sigma_{x}\Omega_{x}^{\lambda}+\sigma_{y}\Omega_{y}^{\lambda})$.
Here, ${\mathbf \sigma}$ is the Pauli matrix. For HHs, we have
\cite{Winkler,Lv_PRB_0603}
\begin{eqnarray}
  \Omega ^{HH}_{x}({\bf K})=2K_{y}(\gamma ^{HH}_{a}K_{y}^{2}
  +\gamma ^{HH}_{b}K_{x}^{2})\ ,
  \label{equ:Omega_HH_1}\\
  \Omega ^{HH}_{y}({\bf K})=2K_{x}(\gamma ^{HH}_{a}K_{x}^{2}
  +\gamma ^{HH}_{b}K_{y}^{2})\ ,
  \label{equ:Omega_HH_2}
\end{eqnarray}
with $\gamma ^{HH}_{a}=E_{z}(\gamma
^{7h7h}_{53}+\gamma^{7h7h}_{54})$, $\gamma ^{HH}_{b}=E_{z}(\gamma
^{7h7h}_{53}-3\gamma ^{7h7h}_{54})$.
For LHs, 
\begin{eqnarray}
  &&\Omega ^{LH}_{x}({\bf K})=2K_{y}(\gamma ^{LH}_{a}K_{y}^{2}+\gamma
  ^{LH}_{b}K_{x}^{2}+\gamma ^{LH}_{c}\langle k_{z}^2\rangle)\ ,
  \label{equ:Omega_LH_1}\\
  &&\hspace{-0.3cm}\Omega ^{LH}_{y}({\bf K})=-2K_{x}(\gamma ^{LH}_{a}K_{x}^{2}+\gamma
  ^{LH}_{b}K_{y}^{2}+\gamma ^{LH}_{c}\langle k_{z}^2\rangle)\ ,
  \label{equ:Omega_LH_2}
\end{eqnarray}
with $\gamma^{LH}_{a}=E_{z}(\gamma
^{6l6l}_{53}+\gamma^{6l6l}_{54})$, $\gamma ^{LH}_{b}=E_{z}(\gamma
^{6l6l}_{53}-3\gamma ^{6l6l}_{54})$, $\gamma ^{LH}_{c}=E_{z}\gamma
^{6l6l}_{52}$. It is noted from these equations that the magnitude
of the Rashba term can be tuned by the external electric field
$E_z$ applied on the sample.\cite{Luo,Nitta} In Eqs.\
\eqref{equ:Omega_HH_1}-\eqref{equ:Omega_LH_2},
$\gamma^{7h7h}_{53}$, $\gamma^{7h7h}_{54}$, $\gamma^{6l6l}_{52}$,
$\gamma^{6l6l}_{53}$, $\gamma^{6l6l}_{54}$ are the Rashba
coefficients.\cite{Winkler} They depend both on the property of
material and QW well width.

Similar to Refs.\ \onlinecite{Cheng_APL_05} and
\onlinecite{Jiang_JAP_06}, by employing the Floquet theorem,
\cite{Shirley} the solution of the Schr\"odinger equation with 
time-dependent Hamiltonian $H_{\lambda}({\bf K},t)$ can be written as 
\begin{eqnarray}
  \nonumber
  \Phi^{\lambda}_{s}({\bf k},t)&=&e^{ -i\{(E^{\lambda}_{\bf k}+E^{\lambda}_{em})t
    -b_{0}{\bf k}\cdot{\bf E}[\cos(\Omega t)-1]-\gamma \sin(2\Omega t) \} }\\
  &&\times e^{-q^{\lambda}_{s}({\bf k})t}
  \sum^{\infty}_{n=-\infty}\phi^{\lambda}_{n,s}({\bf k})e^{in\Omega t}
  \label{solution}
\end{eqnarray}
Here $s={\pm}$ represents the two helix spin branches;
$E^{\lambda}_{\bf k}={\bf k}^2/2m^{\ast}_{\lambda}$ is the kinetic
energy of HHs or LHs;
$E_{em}^{\lambda}=e^2E^2/(4m^{\ast}_{\lambda}\Omega^2)$ is the
energy induced by the RF due to the DFKE;
$b_0=e/(m^{\ast}_{\lambda}\Omega^2)$; $\gamma=E_{em}/(2\Omega)$.
$\phi_{n,s}^{\lambda}({\bf k})=(\phi_{n,s}^{\lambda,\sigma}({\bf
k})) \equiv\genfrac{(}{)}{0pt}{}{\phi_{n,s}^{\lambda,+1} ({\bf
k})}{\phi_{n,s}^{\lambda,-1}(\bf k)}$ in Eq.\ (\ref{solution}) are
the expansion coefficients of the  Floquet states with
$\sigma=1\ (-1)$ representing spin-up $\uparrow$ (-down
$\downarrow$) in the laboratory coordinates (along the $z$ axis).
$q_s({\bf k})$ is the corresponding eigenvalue and can be
determined by
\begin{widetext}
\begin{eqnarray}
  \nonumber
  &&[n\Omega -q_s^{\lambda}({\bf k})]\phi^{\sigma}_{n,s}
  +\big\{[D^{\lambda}_{01}({\bf k})\pm i\sigma D^{\lambda}_{02}({\bf k})]
  +2(e/2\Omega)^2[D^{\lambda}_{21}({\bf k})\pm
  i\sigma D^{\lambda}_{22}({\bf k})]\big\}
  \phi^{-\sigma}_{n,s}\\ &&
  \nonumber
  {}+\big\{i(e/2\Omega)[D^{\lambda}_{11}({\bf k})\pm i\sigma D^{\lambda}_{12}({\bf k})]
  +3i(e/2\Omega)^3[D^{\lambda}_{31}({\bf k})\pm
  i\sigma D^{\lambda}_{32}({\bf k})]\big\}
  (\phi^{-\sigma}_{n+1,s}-\phi^{-\sigma}_{n-1,s}) \\ &&
  {}-(e/2\Omega)^2[D^{\lambda}_{21}({\bf k})\pm i\sigma D^{\lambda}_{22}({\bf k})]
  (\phi^{-\sigma}_{n+2,s}+\phi^{-\sigma}_{n-2,s})
  -i(e/2\Omega)^3[D^{\lambda}_{31}({\bf k})\pm i\sigma D^{\lambda}_{32}({\bf k})]
  (\phi^{-\sigma}_{n+3,s}-\phi^{-\sigma}_{n-3,s})=0\ ,
  \label{equ:eigen}
\end{eqnarray}
where
\begin{align*}
  \hspace{-1cm}
  &D^{HH}_{01}=\gamma_a^{HH} k_y^3+\gamma_b^{HH} k_x^2 k_y\ ,&
  &D^{HH}_{02}=\gamma_a^{HH} k_x^3+\gamma_b^{HH} k_y^2 k_x\ ,\\
  &D^{LH}_{01}=\gamma_a^{LH} k_y^3+\gamma_b^{LH} k_x^2 k_y
  +\gamma_c^{LH} \langle k_{z}^2\rangle k_y\ ,&
  &D^{LH}_{02}=\gamma_a^{LH} k_x^3+\gamma_b^{LH} k_y^2 k_x
  +\gamma_c^{LH} \langle k_{z}^2\rangle k_x\ ,\\
  &D^{HH}_{11}=3\gamma_a^{HH}k_y^2 E_y+\gamma_b^{HH} (k_x^2 E_y
  +2k_xk_yE_x)\ , &
  &D^{HH}_{12}=3\gamma_a^{HH}k_x^2 E_x+\gamma_b^{HH} (k_y^2 E_x
  +2k_yk_xE_y)\ , \\
  &D^{LH}_{11}=3\gamma_a^{LH}k_y^2 E_y+\gamma_b^{LH} (k_x^2 E_y
  +2k_xk_yE_x)+\gamma_c^{LH} \langle k_{z}^2\rangle E_y\ , &
  &D^{\lambda}_{22}=3\gamma_a^{\lambda} k_x E_x^2+\gamma_b^{\lambda}
  (2k_yE_xE_y+k_xE_y^2)\ ,  \\
  &D^{\lambda}_{21}=3\gamma_a^{\lambda} k_y E_y^2+\gamma_b^{\lambda}
  (2k_xE_yE_x+k_yE_x^2)\ , &
  &\hspace{-0.5cm}D^{LH}_{12}=3\gamma_a^{LH}k_x^2 E_x+\gamma_b^{LH} (k_y^2 E_x
  +2k_yk_xE_y)+\gamma_c^{LH} \langle k_{z}^2\rangle E_x\ , \\
  &D^{\lambda}_{31}=\gamma_a^{\lambda} E_y^3+\gamma_b^{\lambda}E_x^2E_y\ ,&
  &D^{\lambda}_{32}=\gamma_a^{\lambda} E_x^3+\gamma_b^{\lambda}E_y^2E_x\ .
\end{align*}
\end{widetext}
 All eigenvalues can be written as
$q_{s,n}=q_{s,0}+n\Omega$ where $q_{s,0}$ is the eigenvalue in the
region $(-\Omega/2,\Omega/2]$. It is evident that $q_{s,n}$ and
$q_{s,0}$ are physically  equivalent. We also find $s=+$ branch
and $s=-$ branch satisfying  the relations:
\begin{eqnarray}
  &&\phi_{n,-}^{\sigma}=-\sigma\phi_{-n,+}^{-\sigma,\ast}\ ,
  \label{equ:symmetry0_1}\\ &&
  q_-({\bf k})=-q_+({\bf k})\ .
  \label{equ:symmetry0_2}
\end{eqnarray}

With the help of Green function, we can calculate the
density of states (DOS) $\rho_{\sigma,\sigma}$ and the density of
spin polarization(DOSP)
$\rho_{\sigma,-\sigma}$,\cite{Cheng_APL_05,Jauho_96}
\begin{eqnarray}
  \nonumber
  \rho_{\sigma_1,\sigma_2}(T,\omega)&=&\frac{1}{2\pi}
  \int^{\infty}_{-\infty}d{\bf k}
  \sum_{s=\pm} \, \sum_{{l_1,l_2 \atop n,m}=-\infty}
    ^{\infty}e^{i(n-m)\Omega T} \\
  \nonumber
  &&\hspace{-0.4cm}
  \times R_{\sigma_1,\sigma_2}(s;n,m;{\bf k})
  J_{l_1}(-2b_0{\bf k}\cdot{\bf E}\sin(\Omega T))\\
  \nonumber
  &&\hspace{-0.4cm}
  \times J_{l_2}(2\gamma \cos(2\Omega T))
  \delta(\omega-[E_{\bf k}+E_{em}\\
  &&\hspace{-0.4cm}
  {}-(l_1+2l_2+n+m)\Omega/2+q_s({\bf k})])\ ,
  \label{equ:Density_Matrix}
\end{eqnarray}
in which $J_n(x)$ is the Bessel function of $n$th order,
$R_{\sigma_1,\sigma_2}(s;n,m;{\bf k}) =({\eta_{\sigma_1}}^{\dag}
\phi_{n,s}({\bf k})) ({\phi_{m,s}}^{\dag}({\bf
k})\eta_{\sigma_2})$ with $\eta_{\sigma}$ standing for the
eigenfunction of $\sigma_z$. It is seen from Eq.\
\eqref{equ:Density_Matrix} that these densities are periodic
functions of $T$ with period $T_0$. The DOSP is nonzero only when
both the RF and the SOC are present. Furthermore, the induced
magnetic moment can be written as
\begin{eqnarray}
  \hspace{-0.5cm}
  \nonumber
  {\bf M}(T)&=&
  \Big( M_x(T),M_y(T),M_z(T) \Big)\\
  \nonumber
  &=&\frac{2g\mu_B}{n_{\uparrow}+n_{\downarrow}}
  \int_{-\infty}^{E_F(T)}\,d\omega
  \Big(\mbox{Re}\,\rho_{\uparrow,\downarrow},
  -\mbox{Im}\,\rho_{\uparrow,\downarrow}, \\ &&
  \frac{1}{2}(\rho_{\uparrow,\uparrow}-\rho_{\downarrow,\downarrow})
  \Big) \ ,
  \label{Magnetic_Moment}
\end{eqnarray}
where the Fermi energy $E_F(T)$ is determined by
$n_{\sigma}=\int_{-\infty}^{E_F(T)}
  \rho_{\sigma,\sigma}(\omega,T)\,d\omega$ where
$n_{\sigma}$ represents the hole concentration. Eq.\
\eqref{Magnetic_Moment} has been simplified by using the fact that
$\rho_{\sigma_1,\sigma_2}=\rho_{\sigma_2,\sigma_1}^{\ast}$. It is
evident that $E_F(T)$ and ${\bf M}(T)$ both oscillate with the
period $T_0$. Due to time reversal symmetry, the DOSP is an odd
function of the time
$\rho_{\sigma,-\sigma}(T,\omega)=-\rho_{\sigma,-\sigma}(-T,\omega)$,
and therefore the DOSP averaged over time reduces to zero.
Besides, the DOS is an even function
$\rho_{\sigma,\sigma}(T,\omega)=\rho_{\sigma,\sigma}(-T,\omega)$
and
$\rho_{\uparrow,\uparrow}(T,\omega)=\rho_{\downarrow,\downarrow}(T,\omega)$,
thus the RF in the $x$-$y$ plane cannot induce magnetic moment
along the $z$ axis. These characters are similar to those of a
2DEG.\cite{Cheng_APL_05}

We numerically solve the eigen-equation Eq.\ \eqref{equ:eigen} and
calculate the DOS and the DOSP through Eq.\
\eqref{equ:Density_Matrix}. One can further obtain the magnetic
moment by using Eq.\ \eqref{Magnetic_Moment}. In the calculation
we choose $a=10$ nm, $E_z=30$ kV/cm. The material parameters of
GaAs are as follows:\cite{Winkler,Landolt} $\gamma_{1}=6.85$,
$\gamma_{2}=2.1$, $\gamma_{3}=2.9$, $\Delta_{0}=0.341$ eV,
$g_{LH}=1.2$, $g_{HH}=3.6$, $m^{\ast}_{LH}=0.0537\ m_0$ and
$m^{\ast}_{HH}=0.171\ m_0$. In order to ensure the validity of the
model which we adopt, we must keep the highest sideband of HH well
separated from the lowest sideband of LH. Moreover, the HH and LH
bands can be splitted by 50 meV by adjusting the applied
strain.\cite{Nordstrom_98} According to these, we choose $E=0.1$
kV/cm, $\Omega=0.1$ THz in the following calculation.

\begin{figure}[htbp]
  \begin{center}
    \includegraphics[width=6.5cm,height=5cm]{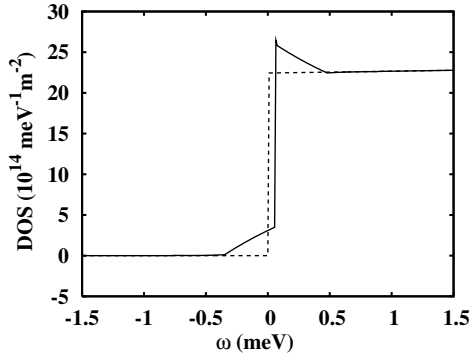}
  \end{center}
  \caption{Time-averaged DOS of HHs under THz field with $E=0.1$
    kV/cm, $\Omega=0.1$ THz (solid curve) and without THz field
    (dashed curve). 
  }
  \label{fig:pw_uu}
\end{figure}

\begin{figure}[htbp]
  \begin{center}
    \includegraphics[width=8cm,height=9cm]{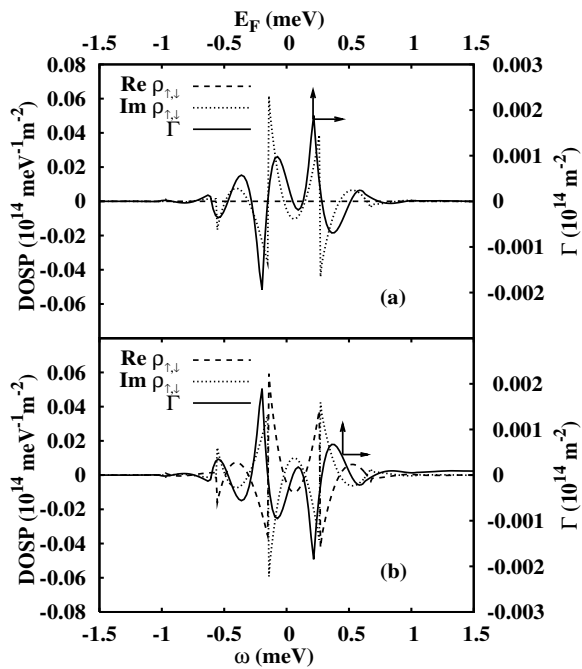}
  \end{center}
  \caption{DOSP of HHs for (a) $\theta_E=0$ and (b) $\theta_E=\pi/4$ at
    $T=T_0/4$ with $E=0.1$ kV/cm, $\Omega=0.1$ THz.
    The real part and imaginary part are plotted as dashed curves and
    dotted curves respectively. The solid curves is $\Gamma$ as
    a function of $E_F$. Note the scale of $E_F$ and $\Gamma$ are on
    the upper and right frame of the figure. 
  }
  \label{fig:pw_ud}
\end{figure}

\begin{figure}[htbp]
  \begin{center}
    \includegraphics[width=6cm,height=5cm]{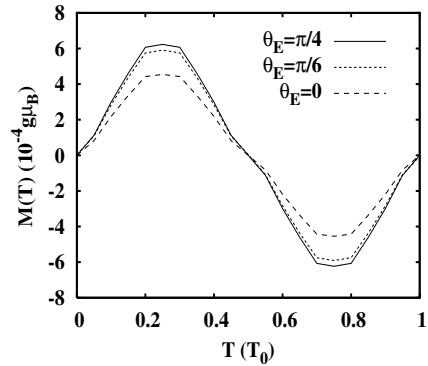}
  \end{center}
  \caption{Magnetic moment $M$ of HHs
    versus  time for $\theta_E=0$, $\theta_E=\pi/6$ and $\theta_E=\pi/4$
    with $E=0.1$ kV/cm, $\Omega=0.1$ THz. The concentration of HH is
    $n_{\uparrow}=n_{\downarrow}=0.7\times 10^{11}$ cm$^{-2}$ and 
    $E_F$ is about 0.35 meV.
  }
  \label{fig:M_all}
\end{figure}

In Fig.\ \ref{fig:pw_uu} we compare the time-averaged DOS with and
without the THz field. Due to DFKE, the main absorption edge has a
blueshift and the DOS becomes finite below the band gap. The DOSP
at $T=T_0/4$ are plotted in Fig.\ \ref{fig:pw_ud} for THz field along two
different directions, (a) $\theta_{E}=0$ and (b) $\theta_{E}=\pi/4$.
It is seen from Fig.\ \ref{fig:pw_ud}(a) that only the
imaginary part of DOSP is finite. From Eq.\ \eqref{Magnetic_Moment},
we can find that the induced magnetic moment is along the $y$ axis.
This is similar to the 2DEG case with Rashba SOC.\cite{Cheng_APL_05}
In Fig.\ \ref{fig:pw_ud}(b), we can see that
$\mbox{Re}\,\rho_{\uparrow,\downarrow}=-\mbox{Im}\,\rho_{\uparrow,\downarrow}$.
Thus the induced magnetic moment is along the $(1,1,0)/\sqrt{2}$
direction, i.e., parallel to the THz field. These results indicate
that the direction of the induced magnetic moment varies with that of
the THz field, which is different from the 2DEG case with Rashba
SOC.\cite{Cheng_APL_05} This is due to the anisotropy of the SOC
Hamiltionian. 

In Fig.\ \ref{fig:M_all}, the magnetic moment of $M$ 
is plotted as function of time for
$\theta_E=0$, $\theta_E=\pi/6$ and $\theta_E=\pi/4$ with $E=0.1$
kV/cm, $\Omega=0.1$ THz. It is noted that that the magnitude of
magnetic moment depends on the direction of the THz field
$\theta_{E}$. The magnetic moment is the smallest for $\theta_E=0$ and
the largest for $\theta_E=\pi/4$. 

In Fig.\ \ref{fig:pw_ud}, we also plotted $\Gamma$ as function of
$E_F$, where
$\Gamma=\int_{-\infty}^{E_F}\mbox{Im}\,\rho_{\uparrow,\downarrow}\,d\omega$.  
It is noted that $\Gamma$ is very small for large enough 
$E_F$, hence $M$ becomes negligible when the concentration of
the hole gas is high. This can be understood as follows:
By interchanging the order of integral, one has
\begin{eqnarray*}
  \int_{-\infty}^{\infty}\rho_{\uparrow,\downarrow}\,d\omega
  &=&\int_{-\infty}^{\infty}\,d{\bf k}
  \sum_{l_1}J_{l_1}(-2b_0{\bf k}\cdot{\bf E}\sin(\Omega T))\\ &&
  \times\sum_{s,m,n}R_{\uparrow,\downarrow}(s;n,m;{\bf k})
  e^{i(n-m)\Omega T}\\ &&
  \times\sum_{l_2}J_{l_2}(2\gamma \cos(2\Omega T)) \ .
\end{eqnarray*}
By virtue of Eqs.\ 
\eqref{equ:symmetry0_1} and \eqref{equ:symmetry0_2}, one gets
$R_{\uparrow,\downarrow}(s;n,m;{\bf k})
=-R_{\uparrow,\downarrow}(-s;-m,-n;{\bf k})$. Thus the terms
of $s=+$ branch compensate those of $s=-$ branch, and
the integral of the DOSP over the whole range $(-\infty,\infty)$ is
zero. On the other hand, the DOSP decays to very small value with
increasing $\omega$ due to the interference caused by the momentum
dependence of the SOC. Hence the contribution to the magnetic moment
at large $\omega$ is negligible. Accordingly, $M$ becomes very
small when $E_F$ is large, i.e., the hole concentration is high. 
Our calculation shows that this is also true for 2DEG with Rashba SOC.

In conclusion, we study the effects of the intense THz field on
2DHG with Rashba SOC. We calculate the DOS and DOSP. We also show that
the a THz magnetic moment can be excited for low hole
concentration. It is noted that the direction of the THz field has a
strong influence on the angle between the induced magnetic moment and
the THz field, as well as on the amplitude of the magnetic moment,
which is quite different from 2DEG with Rashba SOC case. We also point
out that the magnetic moment becomes very small if the hole
concentration is high enough, due to the interference caused by the
momentum dependence of the SOC. This effect also appears in 2DEG.

The author would like to thank M. W. Wu for proposing
the topic as well as the directions during the investigation.
This work was supported by the National Natural Science
Foundation of China under Grant No.\ 10574120, the National
Basic Research Program of China under Grant No.\ 2006CB922005,
the Knowledge Innovation Project of Chinese Academy of Sciences 
and SRFDP. The author would also like to thank
J. H. Jiang for helpful discussions and I. C. da Cunha Lima
for proof reading of this manuscript.


\begin{thebibliography}{0}
\bibitem{Franz} W. Franz, Z. Naturforsch. Teil A {\bf 13}, 481 (1958).
\bibitem{Keldysh} L. V. Keldysh, Sov. Phys. JETP {\bf 34}, 788 (1958).
\bibitem{Jauho_96} A. P. Jauho and K. Johnsen, Phys. Rev. Lett.
  {\bf 76}, 4576 (1996).
\bibitem{Johnsen_98} K. Johnsen and A. P. Jauho, Phys. Rev. B
  {\bf 57}, 8860 (1998).
\bibitem{Nordstrom_98} K. B. Nordstrom, K. Johnsen, S. J. Allen,
  A. -P. Jauho, B. Birnir, J. Kono, and T. Noda, Phys. Rev. Lett. 
  {\bf 81}, 457 (1998).
\bibitem{Cerne_97} J. Cerne, K. Kono, T. Inoshita, M. Sundaram, and
  A. C. Gossard, Appl. Phys. Lett. {\bf 70}, 3543 (1997).
\bibitem{Kono_97} J. Kono, M. Y. Su, T. Inoshita, T. Noda, M. S. Sherwin,
  S. J. Allen, Jr., and H. Sakaki, Phys. Rev. Lett. {\bf 79}, 1758 (1997).
\bibitem{Phillips_99} C. Phillips, M. Y. Su, M. S. Sherwin, J. Ko,  and
  L. Coldren, Appl. Phys. Lett. {\bf 75}, 2728 (1999).
\bibitem{Cheng_APL_05} J. L. Cheng and M. W. Wu, Appl. Phys. Lett.
  {\bf 86}, 032107 (2005).
\bibitem{Jiang_JAP_06} J. H. Jiang, M. Q. Weng, and M. W. Wu,
  J. Appl. Phys. {\bf 100}, 063709 (2006).
\bibitem{Jiang_PRB_07} J. H. Jiang and M. W. Wu, Phys. Rev. B
  {\bf 75}, 035307 (2007).
\bibitem{Bychkov} Y. A. Bychkov and E. Rashba, Sov. Phys. JETP Lett. 
  {\bf 39}, 78 (1984).
\bibitem{Winkler} R. Winkler, {\it Spin-Orbit Coupling Effects in
    Two-Dimensional Electron and Hole Systems} (Springer, Berlin,
  2003).
\bibitem{Lv_PRB_0603} C. L\"u, J. L. Cheng, and M. W. Wu,
  Phys. Rev. B {\bf 73}, 125314 (2006); {\it ibid.} {\bf 71}, 075308 (2005).
\bibitem{Luo} J. Luo, H. Munekata, F. F. Fang, and P. J. Stiles,
  Phys. Rev. B {\bf 41}, 7685 (1990).
\bibitem{Nitta} J. Nitta, T. Akazaki, H. Takayanagi, and T. Enoki,
  Phys. Rev. Lett. {\bf 78}, 1335 (1997).
\bibitem{Shirley} J. H. Shirley, Phys. Rev. {\bf 138}, B979 (1965).
\bibitem{Landolt} {\it Numerial Data and Functional Relationships in
    Science and Technology}, Landolt-B\"ornstein, New Series Vol. 17,
  edited by O. Madelung, M. Schultz, and H. Weiss (Springer, Berlin,
  1982).
\end{thebibliography}
\end{document}